\documentclass{aa}  
\usepackage{natbib}
\bibpunct{(}{)}{;}{a}{}{,} 
\usepackage{graphicx}
\usepackage{hyperref}

\usepackage{txfonts}
\usepackage{}

\newcommand{\SN}{SN2025cbj\,}
\newcommand{\sn}{SN2025cbj}
\newcommand{\IC}{IceCube-250421A\,}

\begin{document}

   \title{The Type IIn SN 2025cbj coincidence with the high-energy neutrino IceCube-250421A}

   \author{S.~Garrappa
          \inst{1}
          \and
          E.~A.~Zimmerman\inst{1}
          \and
          T.~Wasserman\inst{1}
          \and
          E.~O.~Ofek\inst{1}
          \and
          A.~Gal-Yam\inst{1}
          \and
          R.~Konno\inst{1}
          \and
          P.~Chen\inst{2,3,1}
          \and 
          O.~Yaron\inst{1}
          \and 
          S.~Ben-Ami\inst{1}
          \and
          C.~M.~Copperwheat\inst{4}
          \and
          S.~Fainer\inst{1}
          \and 
          A.~Horowicz\inst{1}
          \and
          A.~Humpe\inst{4,5}
          \and
          P.~A.~Mazzali\inst{4,6}
          \and
          D.~Polishook\inst{1}
          \and
          E.~Segre\inst{1}
          \and
          S.~A.~Spitzer\inst{1}      
          }

   \institute{Department of Particle Physics and Astrophysics, Weizmann Institute of Science, 76100 Rehovot, Israel\\
   \email{simone.garrappa@gmail.com}\\
   \email{erezimm@gmail.com}\\
   \email{wasserman.tal@gmail.com}
   \and
   Institute for Advanced Study in Physics, Zhejiang University, Hangzhou 310027, China
   \and
   Institute for Astronomy, School of Physics, Zhejiang University, Hangzhou 310027, China
   \and
   Astrophysics Research Institute, Liverpool John Moores University, IC2, Liverpool Science Park, 146 Brownlow Hill, Liverpool L3 5RF, UK
   \and
   Institute for Applications of Machine Learning and Intelligent Systems (IAMLIS), Munich University of Applied Sciences, Lothstr. 34, D-80335 Munich, Germany
   \and
   Max-Planck-Institut f\"ur Astrophysik, Karl-Schwarzschild Str. 1, D-85748 Garching, Germany}

   \date{Published in A\&A}

 
  \abstract 
   { The origins of astrophysical high‑energy neutrino flux remain uncertain. Core‑collapse supernovae (CCSNe) with strong circumstellar material (CSM) interactions (i.e., Type IIn, a.k.a. SNe IIn) are compelling candidates for explaining efficient hadronic acceleration and neutrino production.}
   {We investigate the possible association between the SNe IIn \SN and the IceCube high‑energy neutrino \IC. We assess whether the observed properties of the SN would enable an appreciable neutrino yield.}
   {We combined rapid optical follow‑ups with LAST and archival ZTF photometry with spectroscopy from LT/SPRAT and MMT/BINOSPEC to characterize the SN’s evolution and CSM interaction. We estimated the explosion and peak times from an early light‑curve fitting and quantified the chance‑coincidence probability with resampling simulations that scramble neutrino right ascensions, while preserving declinations and error contours. Using a simple post-shock‑breakout interaction model in a dense wind, we estimated the expected muon‑neutrino yield for IceCube’s real‑time Bronze stream.}
   {The spectra of \SN obtained after the neutrino epoch exhibit persistent narrow Balmer lines superposed on broad Lorentzian electron‑scattering wings, consistent with a sustained dense CSM interaction. For the multimessenger association, resampling the simulations against the TNS catalog gives a chance‑coincidence p-value  of p $\simeq$ 0.24 for observing k$\geq$1 events (and p $\simeq$ 0.078 against the ZTF-BTS catalog). These values are sensitive to the size of the SN and neutrino samples and do not indicate a statistically significant multimessenger association. A post‑breakout interaction scenario predicts an expected N$_{\nu_{\mu}}$ $\sim$ 10$^{-3}$ events in the IceCube Bronze alert stream over 96 days per this one candidate. We discuss the implications of these numbers and the possible biases that could affect these results.}
   {}


   \maketitle
%

\section{Introduction}
The search for astrophysical neutrino sources has been one of the most challenging pursuits in multimessenger astronomy during the past decade. 
Improvements to neutrino telescopes and multiwavelength programs carried out by the astronomical community have unveiled a plethora of interesting candidates. These sources span a range of objects extending from nearby galaxies \citep{2022Sci...378..538I,2025ApJ...981..103S} to peculiar and powerful active galaxies across cosmological distances \citep{2016NatPh..12..807K,IceCube:2018dnn,2019ApJ...880..103G,2020ApJ...893..162F,2023arXiv230511263B,2024A&A...687A..59G}. Rare systems such as tidal disruption events (TDEs; \citealt{2021NatAs...5..510S}, \citealt{2022PhRvL.128v1101R},\citealt{2024MNRAS.529.2559V}), consisting of an interaction between a star and a massive black hole, have also been suggested as possible neutrino sources. 
The introduction of a real-time alert distribution program by the IceCube South Pole Neutrino Observatory in 2016 \citep{2017APh....92...30A} significantly enhanced  the number of apparent spatial and temporal coincidence cases of rare, high-energy (>100 TeV) neutrinos with astrophysical sources observed at various wavelengths. However, since the detection of a diffuse high-energy neutrino flux in 2013 \citep{2013Sci...342E...1I} and a number of improvements in its characterization \citep{2024PhRvD.110b2001A}, its origins remain a mystery. 

Among the proposed contributors to the diffuse high-energy neutrino flux, supernova (SN) explosions within dense circumstellar material (CSM) represent a compelling source population. These include Type IIn SNe (SNe IIn), which are a subclass of SNe that exhibit strong, long-lasting emission lines, as well as a blue spectral continuum that most likely come from the interaction of the SN ejecta with CSM \citep{1990MNRAS.244..269S,2017hsn..book..403S,2017hsn..book..195G}.  In such systems, shock regions are formed where fast ejecta encounter wind or shells around the progenitor star, driving diffusive shock acceleration of ions up to TeV–PeV energies. The dense CSM environment is an ideal target for those hadrons to produce pions via inelastic $pp$ interactions, which, in turn, decay into high-energy neutrinos and gamma rays. After the shock breaks out into the optically thick wind \citep{2010ApJ...724.1396O}, a collisionless shock is expected to develop and multiTeV neutrinos are expected to be produced efficiently on timescales that range from hours to months (\citealt{2012IAUS..279..274K}, \citealt{2014MNRAS.440.2528M}, \citealt{Murase2018}). 

In addition, recent evidence has revealed that a large fraction of SNe IIn are located within compact distributions of CSM \citep{2021ApJ...912...46B,2023ApJ...952..119B}. The resulting neutrino flux in these large number of sources could account for a significant fraction of the observed $>10$~TeV neutrino background \citep{2025ApJ...978..133W}. Because these dense environments are usually opaque to gamma-ray radiation, high-energy neutrinos are the best messenger to probe cosmic-ray acceleration at TeV-PeV energies. \cite{2022ApJ...929..163P} investigated the possibility that the neutrino TDE candidate AT2019fdr could be a SN IIn instead. 
A stacking analysis using seven years of IceCube data and a sample of $>$1000 core-collapse SNe (CCSNe; including types IIn, IIP, and stripped‑envelope events) found no significant signal above the background \citep{IceCubeSNStack2023}.  The derived 90\% confidence limit, assuming a E$^{-2.5}$ spectrum, constrains the  emission in high-energy neutrinos (10$^{3}$–10$^{5}$ GeV) from SNe IIn to $\lesssim$ 1.3$\times$ 10$^{49}$ {\rm erg}. This corresponds to a contribution of $\lesssim$ 34\% to the diffuse neutrino flux, assuming the E$^{-2.5}$ spectral shape \citep{IceCubeSNStack2023}. 
Although these results do not identify a population of detected neutrino emitters, the limits obtained still leave room for the association of a few sources with single high-energy events. 

On 2012 March 30, a $\sim$2.7$\sigma$ alert from the IceCube optical follow-up program (OFU;\citealt{2009arXiv0909.0631F}) consisting of two $\sim$ TeV neutrinos detected within < 2 s and with a 1.32 deg angular separation was found to be coincident with the SN IIn PTF12csy \citep{2015ApJ...811...52A}. However, the a posteriori significance of this coincidence was estimated only at a 2.2$\sigma$ level. 

\cite{Lu2026} discussed the association of two SN IIn with events from the IceCube realtime stream of track-like high-energy neutrinos. They found a low ($p\sim$ 0.7\%) probability that these two coincidences were observed by chance. We discuss one of these two coincidences, occurring between the neutrino event \IC and \SN, in this work.
Another example is the recent study on the association between an interacting SN Ibn and a high-energy neutrino \citep{2025arXiv250808355S}, which has been estimated to have a probability of $p$ $\sim$ 0.3\% to occur by chance.

In this work, we discuss the association between the SN IIn \SN and the high-energy neutrino \IC detected by IceCube. In Sect. \ref{sec:neutrino_event}, we present the detection of the high-energy neutrino event and how it was found in coincidence with \SN in our real-time follow-up program. In Sect. \ref{sec:observations}, we present the photometric and spectroscopic observations of \SN and explain how we derived useful physical quantities to characterize the SN. In Sect. \ref{sec:discussion}, we discuss the multimessenger coincidence by means of resampling simulations. In Sect. \ref{sec:neutrino_rate}, we estimate an expected neutrino rate for \SN based on the general model from \cite{2012IAUS..279..274K}.

\section{The high-energy neutrino event \IC}\label{sec:neutrino_event}
On 2025 April 21, at 17:06:08.07 UTC, IceCube detected a track-like event with a 48.1\% probability (i.e., signalness) of being astrophysical in origin \citep{2025GCN.40195....1I}. The event belongs to the Bronze alert selection stream, whose average signalness is 30\%. The best-fit reconstructed direction of the neutrino is at RA$_{J2000}$ = 240.91 (+2.62, -4.44) deg, Dec$_{J2000}$ = 28.67 (+1.70,-1.70) deg, with the errors referring to the extension of the 90\% PSF containment. The relatively large errors in the reconstructed direction are due to the fact that the muon track was crossing a short amount of detector volume at one of the detector corners. The event has an estimated energy of 151.43 TeV, and its false alarm rate due to atmospheric background is 1.0254 events per year \citep{2025GCN.40195....1I}.

As part of the Large Array Survey Telescope (LAST) follow-up program for multimessenger alerts and given the favorable observability of the neutrino arrival direction, we automatically triggered observations of the neutrino 90\% error region at 22:19:54 UTC of the same day, 5.23 hours after the IceCube detection. The observations covered about 90\% of the neutrino footprint reported in \cite{2025GCN.40195....1I} by observing a single predefined LAST survey field. The LAST automatic transient-detection pipeline (Konno et al., in prep.) found one candidate transient optical counterpart in the observed neutrino footprint at RA$_{J2000}$ = 242.22 deg, Dec$_{J2000}$ = 27.0 deg \citep{2025GCN.40208....1G}. At the time of the LAST observations, this transient was already known as \SN, a SN candidate classified as Type IIn at redshift z = 0.0675 \citep{2025TNSCR.915....1S}. This transient was discovered by the Zwicky Transient Facility (ZTF, \citealt{2019PASP..131a8002B}) on 2025 February 20 (about 60 days before the neutrino detection), located at $\sim$1.8 deg distance from the IceCube best-fit localization, within the 90\% PSF containment.

Among known neutrino-source candidates, a number of gamma-ray blazars has been reported coincident with \IC in \textit{Fermi}-LAT realtime follow-up observations \citep{2025GCN.40242....1P}. They report the coincidence of five known gamma-ray sources. Among these, four are associated with blazars in the 4FGL-DR4 catalog \citep{2023arXiv230712546B} and one is unassociated. In addition, a new gamma-ray source is reported for the first time and tentatively spatially associated to a known blazar detected in radio bands. However, none of these sources have been significantly detected in \textit{Fermi}-LAT data integrated over timescales of days to months prior the neutrino detection. Despite the so-called gamma ray and neutrino connection in blazars is still not clear (e.g., \citealt{2024A&A...687A..59G}), this discourages the association of any of these sources with \IC in the hypothesis unlike previous candidates (e.g., \citealt{IceCube:2018dnn}, \citealt{2019ApJ...880..103G}). In addition, we searched for TDE events in the Transient Name Server\footnote{\url{https://www.wis-tns.org/}} (TNS) and the ZTF Bright Transients Survey \citep{ZTF_BTS_paper1,ZTF_BTS_paper2,ZTF_BTS_paper3} and found no known TDE source coincident with the \IC footprint.\\

\section{Observations of \SN}\label{sec:observations}

\subsection{Photometry}
Here, we  describe the photometric measurements of \SN. We also derive the peak and explosion time from the light curves available from various surveys.

\subsubsection{LAST photometry}

We observed the field of \IC with LAST during its commissioning phase. The LAST system consists of an array of optical telescopes located at the Weizmann Astrophysical Observatory\footnote{\url{https://www.weizmann.ac.il/wao/}} in the Israeli Negev desert, designed to study the variable and transient sky (\citealt{2020PASP..132l5004O},\citealt{2023PASP..135f5001O}, \citealt{2023PASP..135h5002B}). The first phase of the LAST system will consist of 72 Rowe-Ackermann Schmidt Astrograph $f$/2.2 telescopes with 28 cm diameter (40 already operative). Each telescope provides a field of view of 3.3 deg $\times$ 2.2 deg and a plate scale of 1.25 arcsec/pixel on camera. LAST observes in clear filter and the typical limiting magnitude (AB) of the system is 19.6 (21.0) in 20 s (20x20 s) exposures. The LAST data reduction pipeline and its products are described in \cite{2023PASP..135l4502O}, while the image subtraction and transient detection pipeline is discussed in Konno et al. (in prep.).

On the night of  April 21, 2025, the observations of the \IC field were carried out under bright moon conditions, with an average limiting magnitude of 19.70 (AB) over a total of 19 epochs \citep{2025GCN.40208....1G}. The average magnitude of \SN during the 19 epochs was m$_{AB}$ = 18.53 $\pm$ 0.07 and no significant variability was observed. LAST monitored the source for nine additional nights. In addition, prior to the neutrino detection, LAST serendipitously observed the field containing \SN during sky-survey operations on  April 4, 2025 (i.e., 17 days before the neutrino's arrival). On this night, the source was detected at m$_{AB}$ = 18.52 $\pm$ 0.06 (statistical error only), comparable with the brightness at the neutrino arrival. The total number of visits (20x20s exposures) observed by LAST for this object is 128. The average flux during each night is shown in the light curve of Fig. \ref{fig:optical_lc} and listed in Table \ref{tab:LAST_photometry}. The photometric calibration of each image is obtained with the method described in \cite{2025A&A...699A..50G}.

\begin{table}[h!]
    \centering
    \caption{Nightly average of LAST photometric observations of \SN.}
    \begin{tabular}{|lll|}
    \hline
             MJD &  Mag (AB) &  Mag Err. \\
    \hline
    60767.989 &   18.428 &        0.039 \\
    60768.995 &   18.436 &        0.042 \\
    60769.974 &   18.488 &        0.043 \\
    60786.930 &   18.530 &        0.016 \\
    60794.899 &   18.600 &        0.013 \\
    \hline
    \end{tabular}

    \tablefoot{The errors are obtained from error propagation of statistical errors in each epoch. Table \ref{tab:LAST_photometry} is published in its entirety in electronic format at the CDS.}
    \label{tab:LAST_photometry}
\end{table}

\subsubsection{ZTF photometry}\label{sec:photometry}

We used photometric measurements from ZTF \citep{2019PASP..131a8003M} retrieved from the Automatic Learning for the Rapid Classification of Events (ALeRCE) alert broker \citep{2021AJ....161..242F}. The transient \SN (a.k.a. ZTF25aagbrpb) has been observed by ZTF with $g$ and $r$ filters during sky-survey operations. In Fig. \ref{fig:optical_lc}, we show the ZTF light curves. We list the measurements in Table \ref{tab:ztf_photometry}. 

\begin{table}[h!]\label{tab:ztf_photometry}
    \centering
    \caption{ZTF photometric observations of \SN.}
    \begin{tabular}{|l l l l|}
    \hline
             MJD &  Mag &  Mag Err. &  Filter \\
    \hline
    60725.505 & 19.19 &    0.10 &    ZTF-$r$ \\
    60727.489 & 19.09 &    0.09 &    ZTF-$r$ \\
    60727.530 & 18.90 &    0.07 &    ZTF-$g$ \\
    60729.433 & 18.69 &    0.08 &    ZTF-$g$ \\
    60729.494 & 18.87 &    0.09 &    ZTF-$r$ \\
    \hline
    \end{tabular}

    \tablefoot{Table \ref{tab:ZTF_photometry} is published in its entirety in electronic format at the CDS.}
    \label{tab:ZTF_photometry}
\end{table}

\subsubsection{Other multiwavelength follow-up observations}
\begin{figure*}
\sidecaption
    \includegraphics[width=12cm]{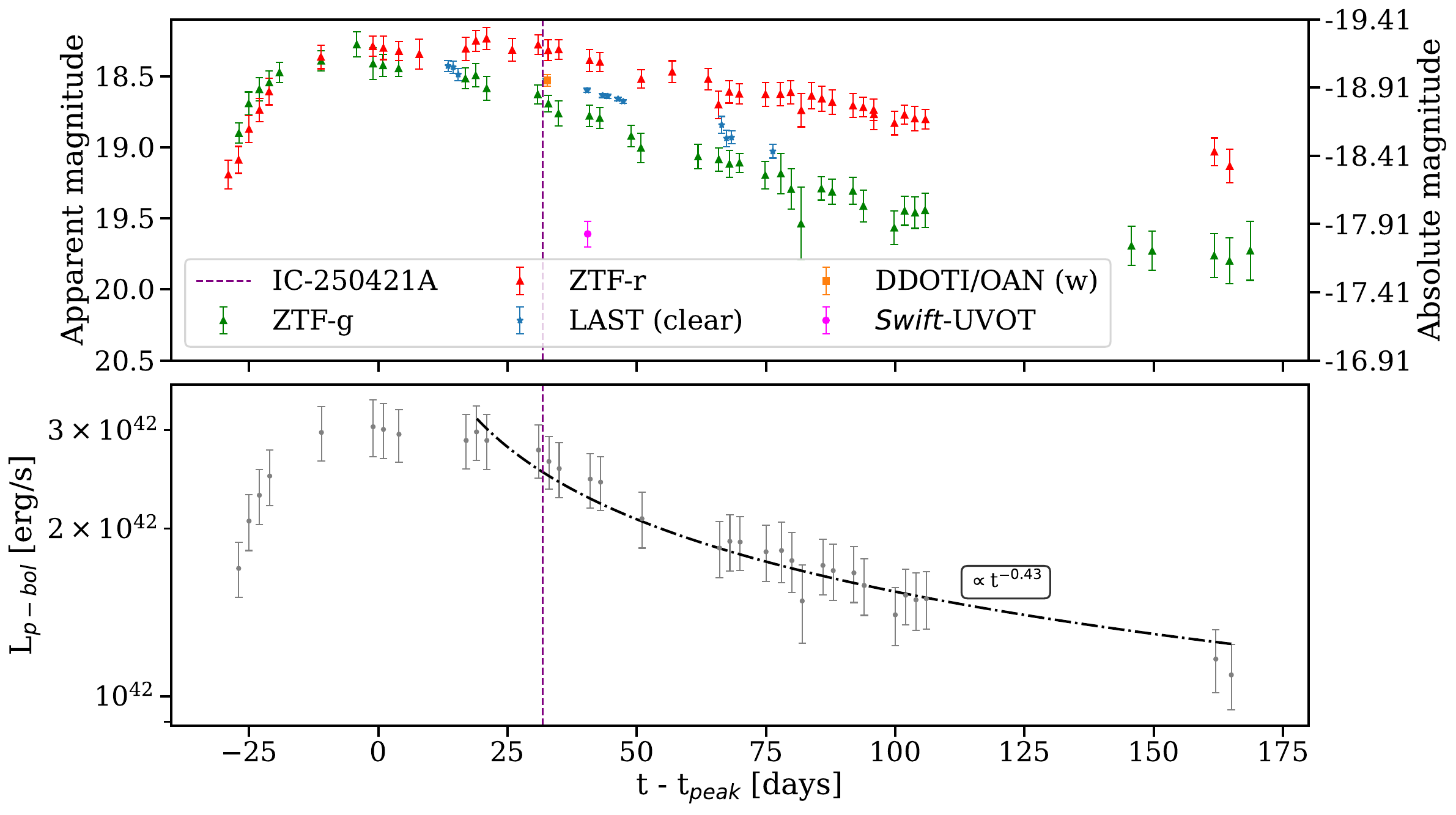}
        \caption{Upper: Photometric light curves from LAST (blue), ZTF-$g$ (green), ZTF-$r$ (red), DDOTI/OAN (orange) with temporal x-axis centered on the fitted peak time at MJD 60754.5. The magenta marker is the \textit{Swift}-UVOT measurement in the UVW1 band. The dashed purple vertical line marks the arrival of \IC. Lower: Pseudo-bolometric light curve in the (4700-6231 $\AA$) range. The dot-dashed line shows the power-law fit of the early light curve decay after the peak ($\propto$ t$^{\alpha}$, $\alpha$ = -0.43).
        \label{fig:optical_lc}}
\end{figure*}

After the LAST report of the possible multimessenger coincidence, the source was observed by DDOTI/OAN \citep{2025GCN.40211....1A,2016SPIE.9910E..0GW}, which reported a flux level in the w (white) band of m$_{w}$ = 18.53 $\pm$ 0.04, consistent with LAST measurements (included in Fig. \ref{fig:optical_lc}).\\
The X-ray Telescope \citep[XRT;][]{2005SSRv..120..165B} on board the \textit{Neil Gehrels Swift Observatory} (\textit{Swift}) observed \SN in a single target of opportunity (PI Veres) visit after the detection of \IC. The observations were carried out on 2025 April 30 at 12:46 UTC for a total exposure of 2259 s. We used the \textit{Swift}-XRT data products generator to reduce the data \citep{2020ApJS..247...54E}. No significant source was detected in \textit{Swift}-XRT observations and we calculated a 3-$\sigma$ upper limit in the 0.3-10 keV range of <1.2$\times$10$^{-13}$ erg cm$^{-2}$ s$^{-1}$ (unabsorbed flux, using N$_{H}$ = 4.1$\times$10$^{20}$). This corresponds to an upper limit on the luminosity of $L_{XRT}\leq$ 1.4$\times$10$^{42}$ erg s$^{-1}$. The source was also observed during the same time with the Ultra-Violet/Optical Telescope \citep[UVOT;][]{2005SSRv..120...95R} on board \textit{Swift} for a total of 2566.191 s exposure, using the UVW1 filter. We detected the source in UVOT data with apparent magnitude of m$_{W1}$ = 19.61 $\pm$ 0.09, corresponding to $f$(2634$\AA$) = (1.561 $\pm$ 0.129)$\times$10$^{-2}$ mJy.

As part of the realtime follow-up program of the Large Area Telescope \citep[LAT;][]{2009ApJ...697.1071A} on board the \textit{Fermi} satellite, a search for gamma-ray emission at the localization of \SN was conducted in light of the multimessenger coincidence reported by LAST. Integrating observations of one month before the neutrino detection, no significant gamma-ray emission was detected \citep{2025GCN.40242....1P}. An upper limit on the flux was reported for a point source with power-law spectrum of index 2.0 at < 1.13$\times$10$^{-8}$ ph cm$^{-2}$ s$^{-1}$ (95\% confidence limit). This corresponds to an upper limit on the gamma-ray luminosity in the 100 MeV - 1 TeV band of L$_{\gamma}$ $\lesssim$ 1.8$\times$10$^{44}$ erg s$^{-1}$. These additional multiwavelength observations are summarized in Table \ref{tab:mwl_obs}.

\begin{table*}
    \centering
    \caption{Multiwavelength observations of \SN.}
    \begin{tabular}{|llll|}
    \hline
        Survey & Band & Flux & MJD \\
        \hline
        DDOTI/OAN & Optical (w) & 18.53 $\pm$ 0.04 & 60794.9 \\
        \textit{Swift}-UVOT& UVW1  & 19.61 $\pm$ 0.09 mag & 60795.5 \\
        \textit{Swift}-XRT & 0.3-10 keV & <1.2$\times$10$^{-13}$ erg cm$^{-2}$ s$^{-1}$  & 60795.5 \\
        \textit{Fermi}-LAT& 100 MeV - 1 TeV & < 1.13$\times$10$^{-8}$ ph cm$^{-2}$ s$^{-1}$ & 60756.7-60786.7 \\
    \hline
    \end{tabular}
    \tablefoot{This is a collection of results from the multiwavelength follow-up campaign after the detection of \IC.}
    
    \label{tab:mwl_obs}
\end{table*}

\subsection{Optical spectroscopy and classification}\label{sec:spectroscopy}
\SN was classified as a SN IIn  on March 5 2025 \citep{classification_report} based on a spectrum taken by the ``SED'' machine \citep[SEDM;][]{2012SPIE.8446E..86B,2018PASP..130c5003B} mounted on the P60 telescope in the Palomar observatory. The spectrum and classification were reported to the TNS and showed persistent narrow Balmer emission lines indicating long-lasting CSM interaction; thus it was classified as Type IIn.

\begin{figure}[h]
    \centering
    \includegraphics[width=0.5\textwidth]{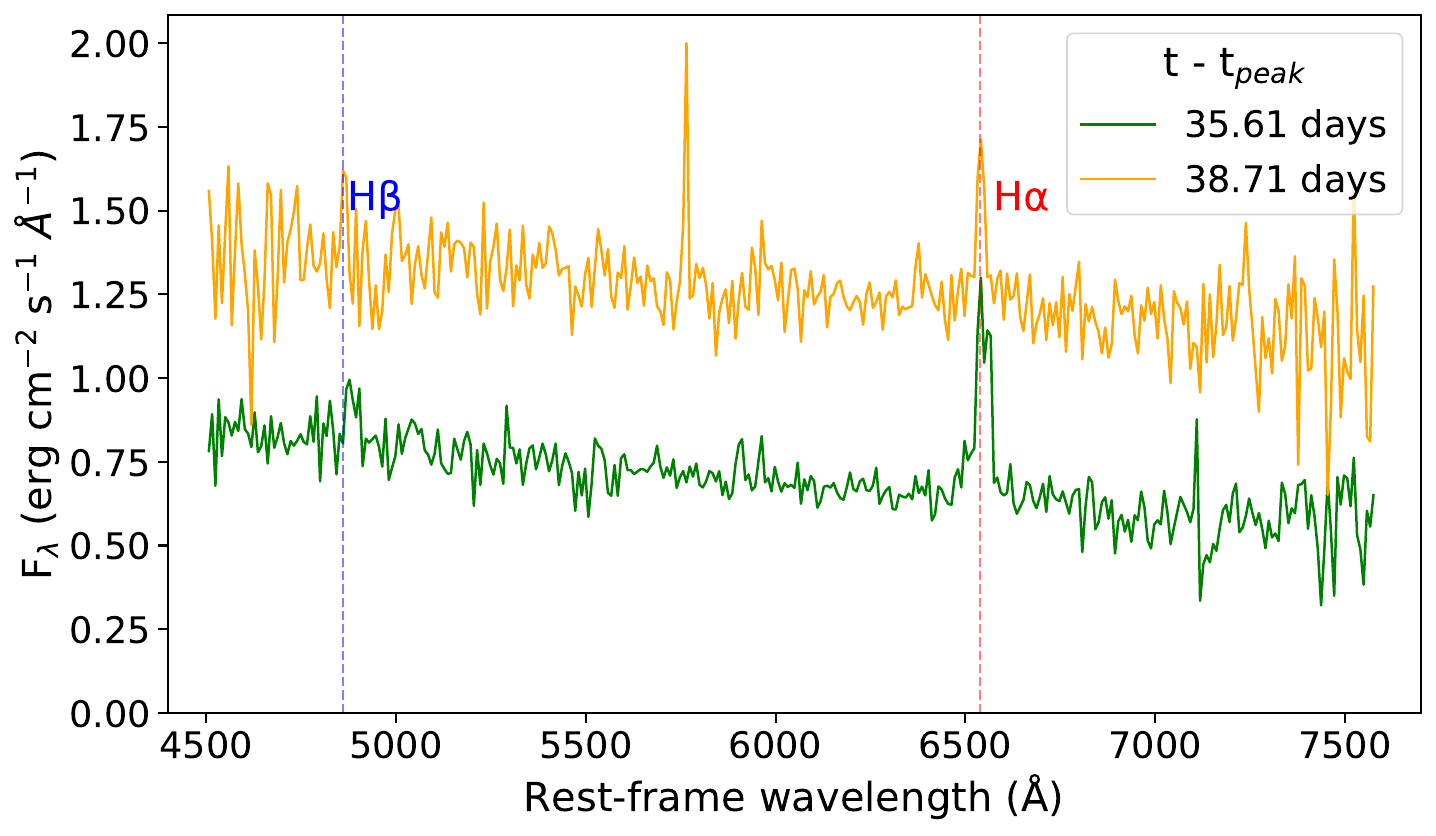}
    \caption{Liverpool telescope spectra of \sn. A narrow H$\alpha$ component can be identified in both epochs, below a strong galactic H $\alpha$ narrow feature. However, the lower resolution does not allow accurate velocity measurements or a positive identification of the H$\beta$ line. We note that the narrow single pixel component in the 22.91 day spectrum is the result of instrumental noise.}.
    \label{fig:LT_Spec}
\end{figure}

Following the identification of the \IC/\SN coincidence, we initiated a spectroscopic follow-up to confirm the SN was still interacting with CSM, which would allow a CSM-driven neutrino production channel. We initially obtained two spectra with the robotic SPRAT spectrograph \citep{2014SPIE.9147E..8HP} mounted on the 2-meter Liverpool Telescope (LT) at the Roque de los Muchachos Observatory on La Palma (see Fig. \ref{fig:LT_Spec}). The spectra were obtained on  April 25 and 28, 2025 and were reduced by the automatic SPRAT pipeline. We identified narrow H$\alpha$ components in both spectra. However, due to the low brightness of \SN, we could not identify any other lines and, thus, we were not able to positively determine whether the narrow H$\alpha$ emission was from the host galaxy, rather than from CSM interaction.

Subsequently, we obtained a third, higher resolution ($R\sim1340$), spectrum using the BINOSPEC \citep{Fabricant2019} spectrograph mounted on the 6.5-meter Multiple Mirror Telescope (MMT) at the MMT Observatory, Arizona on 2025 May 16. We present this spectrum in Fig. \ref{fig:MMT_Spec}. The observation consisted of 3 $\times$ 450 s exposures. The data were acquired with a grating of 270 lines mm$^{-1}$ and a 1.0\,$''$ slit mask. The processed images, after the bias subtraction and flat-fielding using the Binospec pipeline \citep{Kansky2019}, are downloaded from the MMTO queue observation data archive. The spectra are reduced with IRAF, including cosmic-ray removal, wavelength calibration (using arc lamp frames taken immediately after the target observation), and relative flux calibration with an archived spectroscopic standard observation.\\
The MMT spectrum shows clear signs of CSM interaction, including narrow H$\alpha$ and H$\beta$ emission lines, superimposed on broader Lorentzian-shaped electron scattering wings, which are common in interacting SNe \citep{Huang2018}. Interestingly, the H$\alpha$ Lorentzian shape is very asymmetric, with its blue wing being much broader than its red wing. This is also seen to a lesser extent in H$\beta$. This effect has been observed in other interacting SNe such as SN\,2013L \citep{Andrews2017} and SN\,2010jl \citep{Smith2012,Ofek2014,Fransson2014} and indicates asymmetries in the CSM. However, since we only obtained a single epoch of medium-resolution spectrum, we cannot compare its evolution to that of other IIn SNe. Regardless, the consistency of narrow emission lines well after the neutrino detection confirms the SN had significant interaction at that time. All the spectra will be available on WISeREP \citep{2012PASP..124..668Y} after the journal publication.

\subsubsection{Narrow-line velocity}
\begin{figure*}
    \centering
    \includegraphics[width=17cm]{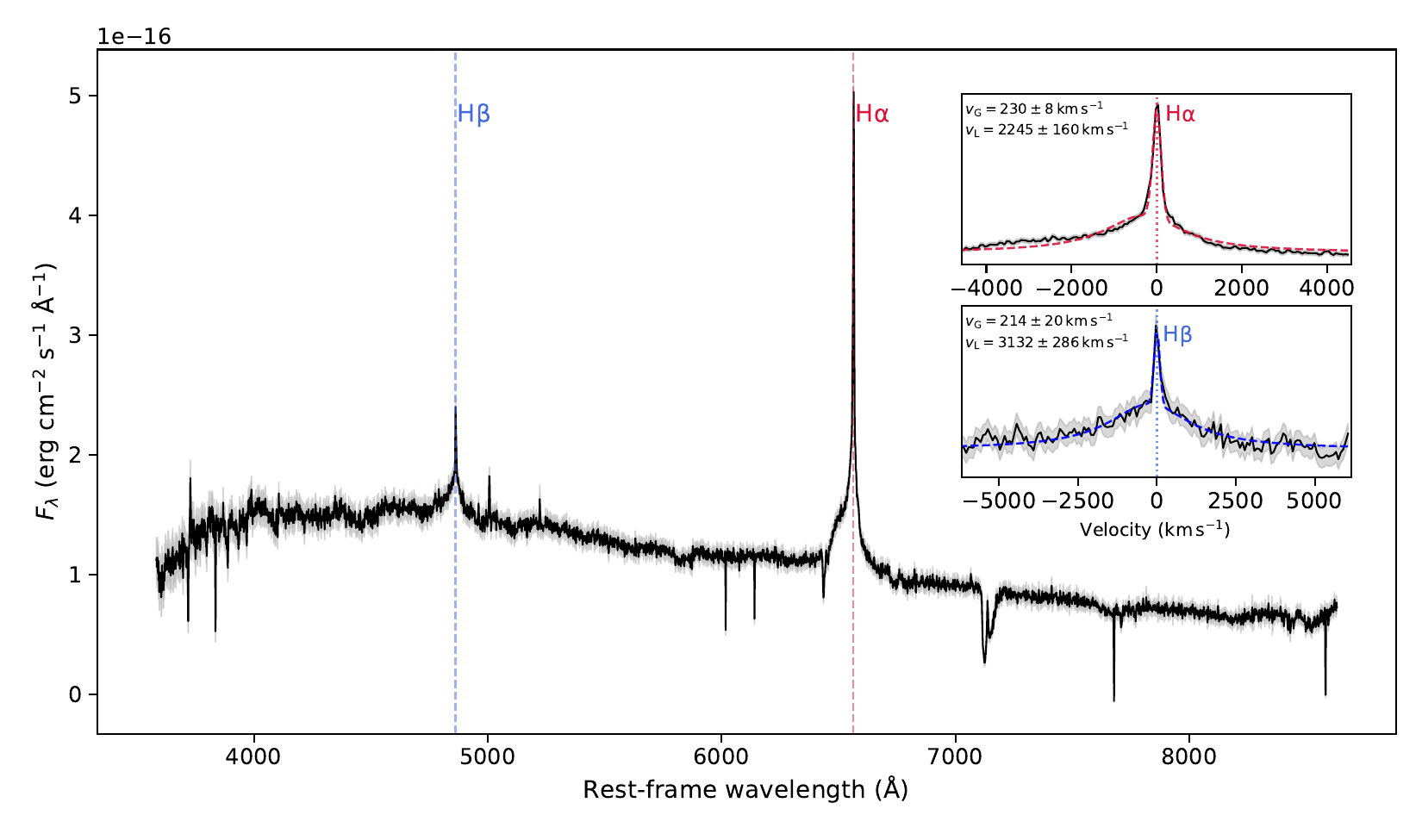}
    \caption{MMT spectrum of \sn $\sim$56 days after the SN peak brightness. Zero velocity H$\alpha$ and H$\beta$ are marked with red and blue striped vertical lines, respectively. Zoom-ins on the H$\alpha$ and $\beta$ profiles with fits to a Lorentzian plus Gaussian shape are added, showing a good fit aside from the blue-wing asymmetry in H$\alpha$. The best-fit velocities (i.e., line FWHM) for both the Gaussian, $v_{G}$, and Lorentzian, $v_{L}$, components are presented next to the fits.}
    \label{fig:MMT_Spec}
\end{figure*}

To measure the velocity of the narrow lines, we fit the Balmer lines in the MMT spectrum using the sum of a Lorentzian and a Gaussian profile, with the Lorentzian component representing the electron-scattering wings and the Gaussian the emission line. We then treated the full width half maximum (FWHM) of the Gaussian component as the bulk velocity of the emission line. We find both lines fit a velocity of $v\sim230\rm\,km\,s^{-1}$. However, this measurement should be treated as an upper limit, as the MMT spectral resolving power at the 270 lines/mm configuration is $R\sim1340$, which corresponds to a similar velocity limit of $\Delta v\sim230\rm\,km\,s^{-1}$. Regardless, the overall profile, including the much broader Lorentzian base, fits the lines well, indicating it is of CSM origin and not host galaxy emission lines. We also note that the Lorentzian line wings are too narrow to be of photospheric origin at $v\sim3000\,\rm km\,s^{-1}$, while typical SN photospheric velocities are closer to $10^{4}\,\,\rm km\,s^{-1}$ \citep{Huang2018}, confirming the Lorentzian wings are indeed of CSM electron scattering origin. We present these fits as insets in Fig. \ref{fig:MMT_Spec}.

\subsection{Pseudo-bolometric light curve}
We calculated a pseudo-bolometric optical light curve (4770-6231 $\AA$) using observations in the ZTF-$r$ and ZTF-$g$ bands, which are the measurements that cover most of the light curve. Since ZTF observations are typically made with a single filter per night, we rebinned the light curve with a bin width of two days and considered only those bins containing at least one ZTF-$r$ and one ZTF-$g$ observation. The pseudo-bolometric luminosities were computed by fitting and then integrating a simple power-law model between the effective wavelengths of each band ($\lambda_{g}$ = 4770 $\AA$, $\lambda_{r}$ = 6231 $\AA$) and the pairs of fluxes in each two-day bin (F$_{g}$,F$_{r}$). The errors were derived via error propagation and assuming a 5\% error on the measured distance.
The pseudo-bolometric light curve is shown in the bottom panel of Fig. \ref{fig:optical_lc}, the luminosity values are integrated only between the $r$ and $g$ bands, and the errors are statistical only. These values can be interpreted as lower limits on the bolometric luminosity.

\subsection{Explosion and peak time}
To estimate the explosion time (i.e., the time of first light), we used the earliest available photometric data for \SN, which is the ZTF-$r$ data. The first $r$-band detection of $m_{g}=19.19\pm0.1$ was registered on 2025 February 19 at 12:07:02 UTC ($\rm MJD = 60725.5$), following a nondetection on  February 17, 2025 at 09:50:29 UTC ($\rm MJD = 60723.41$) with a 5$\sigma$ limiting magnitude of $m_{r}>19.5$. We fit a broken power-law of $f=a\times(t-t_{\rm exp})^{n}$, where $a$ is a fitting constant and $t_{\rm exp}$ is the explosion time, to the early $r$-band curve in flux space, while fixing the flux to $0$ where $t<t_{\rm exp}$. We fit all data up to the g-band maximum. We find the best fit to be a power law of $n=0.344\pm0.056$ and $t_{\rm exp}= 60723.21\pm1.22$ (best fit at February 17, 2025 at 05:14:17 UTC). While this result may seem contradictory to the original nondetection, preceding it by $\sim 4$ hours, the error bar for our fit is quite large, and the nondetection is not very deep, indicating the explosion time could indeed precede it. We also note that SNe interacting with CSM have been shown to have a complex rise, which could change its trend very early, further increasing the uncertainty of the explosion time \cite[as seen in the case of, e.g., SN\,2023ixf;][]{Li2024,Zimmerman2024}.

We estimated the peak time of the ZTF light curve by fitting a simple Gaussian profile for both the ZTF-$g$ and ZTF-$r$ light curves. We find a best-fit value of t$_{peak,g}$ = 60754.5 $\pm$ 1.4 MJD and t$_{peak,r}$ = 60770.3 $\pm$ 3.7 MJD for the g and r bands, respectively. In the rest of this work, we use the value of t$_{peak,g}$ = 60754.5 $\pm$ 1.4 MJD, which is better constrained by the observations.

\subsection{Early light-curve decay}
We fit the early decay of the pseudo-bolometric light curve after the peak time (MJD 60754.5) with a power-law function ($\propto$ t$^{\alpha}$). We find the best-fit value at $\alpha$ = $-$0.43 $\pm$ 0.04. This value is consistent with expectations from interactions with massive CSM as well as with observations (e.g., SN2010jl, \citealt{2012ApJ...759..108S}, \citealt{Ofek2014}). This supports the hypothesis of strong CSM interaction in \SN. The power-law fit of the early decay after the peak is shown in the bottom panel of Fig. \ref{fig:optical_lc}.\\
We also fit the decaying part of the light curve with an exponential profile described by $L = L_{0}\,e^{-t/\tau_d}$, where t is the time since the peak and $\tau$ is the exponential timescale. We find the best-fit exponential timescale of the decay at $\tau_{d}$ = 142.43 $\pm$ 11.28 day. Although with a large error, this timescale is longer than  the one expected from typical $^{56}$Co decay ($\sim$ 111 day). This longer decay timescale is similar to that observed in peculiar SNe such as 2010ij \citep{Ofek2014,Fransson2014,2016MNRAS.456.2622J}. It supports the interpretation that the light curve of \SN is powered by interaction of the SN shock with the CSM.

\section{Estimating the multimessenger coincidence}\label{sec:discussion}

We investigated the possible association between \SN and \IC by estimating the probability of chance coincidence between the two detections. To do so, we considered a collection of IceCube high-energy track-like events, following the one used in \cite{2024A&A...687A..59G}. This collection includes archival and real-time events from the ICECAT-1 catalog (from 2011 to 2020, \citealt{2024yCat..22690025A}) and real-time events distributed through the GCN network from 2016 until the event IceCube-250708A. The initial neutrino sample consists of 391 track-like events. 
For the SN sample, we collected all transients from two catalogs:  Transient Name Server (TNS, \citealt{2021AAS...23742305G}) and  ZTF Bright Transients Survey \citep{ZTF_BTS_paper1,ZTF_BTS_paper2,ZTF_BTS_paper3}. From TNS, we selected 491 transients classified as a Type IIn SN discovered between 2013 May 9 and 2025 July 8. The ZTF-BTS catalog contains a smaller sample of 166 sources discovered in the same time period. All objects in the ZTF-BTS sample are also found in the TNS sample. We  refer to the two statistical tests henceforth as T$_{tns}$ (against the TNS sample) and T$_{bts}$ (against the ZTF-BTS sample).

The purpose of the statistical tests described in this section is to estimate the chance coincidence probability of observing one or more neutrino-and-SN-IIn coincidences within $\delta$t = 61 days from the SN-IIn discovery date (i.e., the time delay observed between the \SN discovery and the detection of \IC), since we would expect the neutrinos to be produced in the interaction between the SN ejecta and the CSM. To homogenize the SN and neutrino samples, additional selection cuts are required:

\begin{itemize}
    \item  Selecting neutrino events detected from 2013 May 09 (the discovery time of the first SN-IIn in the TNS sample) for T$_{tns}$;
    \item Selecting neutrino events detected from 2018 August 14 (the discovery time of the first SN-IIn in the ZTF-BTS sample) for T$_{bts}$;
    \item Selecting neutrinos and SN-IIn at high galactic latitudes ($|b| > $20);
    \item Selecting neutrino events with a 90\% PSF containment that is not larger than the one of \IC (21.06 sq. deg).
    
\end{itemize}
The number of SNe and neutrino alerts for the final samples used in T$_{tns}$ and T$_{bts}$ are listed in Table \ref{tab:pvalues}. We simulated 10$^{5}$ samples of neutrino alerts from the aforementioned list of IceCube neutrinos by scrambling their right ascensions while keeping their declinations and 90$\%$ error regions fixed. This preserves the specific distribution of IceCube alerts as a function of the arrival declination, which reflects the detector sensitivity across the sky. By repeating the search for coincident SNe IIn for each of the simulated samples within 61 days from their discovery date, we can empirically determine the spatial and temporal chance coincidence. We define a coincidence as an instance when there is at least one SN that is spatially coincident with the 90\% error region of a neutrino alert, with the neutrino detection within a time $\delta$t from the SN discovery. Neutrinos coincident with multiple SNe are treated as a single coincidence. 

\subsection{Null hypothesis: the probability for random coincidences}
From the statistical test T$_{tns}$, described in Sect. \ref{sec:discussion}, we find that the distribution of spatial and temporal coincidences with SNe IIn in the background simulations follows a Poissonian distribution with best-fit $\lambda$ = 0.273. Therefore, the probability of observing one or more coincidences for neutrinos with a reconstruction better than \IC in T$_{tns}$ is $p(k\geq1)$ = 0.24. From the statistical test T$_{bts}$, we find that the number of spatial and temporal coincidences in the background simulations follows a Poissonian distribution with best-fit $\lambda$ = 0.082, which gives $p(k\geq1)$ = 0.078. The results are summarized in Table \ref{tab:pvalues}. These $p$-values are both consistent with the background hypothesis (i.e., the coincidence is observed only by chance). 
Regarding the two different SN IIn samples, the TNS database is a richer resource compared to the ZTF-BTS catalog because it includes contributions by a larger number of survey facilities. 

\begin{table}
    \centering
    \caption{Summary of the sample sizes for SN-IIn and neutrino alerts.}
    \begin{tabular}{|l l l l l|}
    \hline
        Test&SN catalog & SN-IIn  & IceCube alerts  & p-value \\
        \hline
        T$_{tns}$ &TNS& 405 & 183 & 0.24 \\
        
         T$_{bts}$& ZTF-BTS& 138 & 134 & 0.078 \\
         \hline
    \end{tabular}
    \tablefoot{Summary from the TNS and ZTF-BTS samples, the number of IceCube alerts and resulting p-values obtained in each statistical test following the selection described in Sect. \ref{sec:discussion}. }
    
    \label{tab:pvalues}
\end{table}

\subsection{Coincidences between SNe IIn and IceCube high-energy track events}\label{sec:broad_sample_SN}
In the previous section, we discuss the statistical significance of the SN-neutrino coincidence we are focusing on in this work. The purpose of this section is to have a more general view on similar coincidences that can be found in archival data. Since SNe are transient sources, it is key to define a temporal timescale in which to define the coincidence beside the spatial match. As we discuss in more detail in Sect. \ref{sec:neutrino_rate}, in typical SNe-IIn interacting with CSM, a reasonable order of magnitude for the timescale of efficient neutrino production is of $\mathcal{O}$(100) days. We ran the spatial and temporal search between the full sample of 343 IceCube alerts and 491 SN IIn from the TNS catalog. We list all the coincidences in Table \ref{tab:all_coincidences}. There are four high-energy neutrino alerts which are coincident with a SN IIn in TNS. One of these, IceCube-240307A, is coincident with two SNe IIn: SN2024bcm and SN2024cui, with $\delta$t = 39.8 and 20.5 days, respectively. At least one of these two coincidences is of random nature, given the very large error region of the neutrino localization (>700 sq. deg). We noticed that all of these spatial coincidences are observed with poorly reconstructed neutrino events, with \IC being the case with both the smallest error area (21.1 sq. deg) as well as the smallest angular separation between the SN and IceCube best-fit localization ($\delta \theta$ = 1.8 deg). We repeated the simulations described in Sect. \ref{sec:discussion} and we found that the probability of observing four or more random coincidences between the IceCube alerts and the TNS sample is $p(k \geq 4)$ = 0.25. 

When we repeated the same matching between the IceCube alerts and the ZTF-BTS sample, we found only three coincidences, as the two objects SN2021jmm and SN2024bcm are not present in the ZTF-BTS catalog. The probability of observing three or more random coincidences between the IceCube alerts and the TNS sample is $p(k \geq 3)$ = 0.07. These $p$-values are consistent with the hypothesis that these coincidences are random in nature.\\

As a final control test, we used samples of SNe Ia from the TNS and the BTS catalogs that are not expected to be high-energy neutrino emitters \citep{Murase2018}. Repeating T$_{tns}$ and T$_{bts}$ with $\delta$t = 100 days, we found 34 (15) neutrino alerts coincident with SNe Ia, over a sample of 11204 (6650) SNe. We estimated a chance coincidence of $p_{Ia,\mathrm{T_{tns}}}$ (k$\geq$34) = 0.46 and $p_{Ia,\mathrm{T_{bts}}}$ (k$\geq$15) = 0.55, which confirms the robustness of our statistical test.

\begin{table}[h!]
\centering
\caption{IceCube alerts coincident with SNe  IIn  from TNS.}
\begin{tabular}{|l l l l l l|}
\hline
IC name & Area 90\% & SN name & $\delta\theta$ & z & $\delta t$ \\
& [sq. deg]& & [deg] & & [day] \\

\hline
210503A & 102.6 & 2021jmm & 5.8 & 0.19 & 72.8 \\
231027A & 23.6 & 2023syz & 2.1 & 0.037 & 37.7 \\
240307A & 721.2 & 2024bcm & 4.1 & 0.143 & 39.8 \\
240307A & 721.2 & 2024cui & 13.3 & 0.1 & 20.5 \\
250421A & 21.1 & 2025cbj & 1.8 & 0.0675 & 60.5 \\

\hline
\end{tabular}\label{tab:all_coincidences}

\tablefoot{The area is the published 90\% containment error of the neutrino reconstruction, $\delta \theta$ the angular distance between the neutrino and the SN, z the SN redshift, and $\delta t$ the temporal distance between the SN discovery and the neutrino detection.}

\end{table}

\subsection{Comparison with previous works}
A recent work by \cite{Lu2026} investigates the coincidence between SNe IIn and IceCube high-energy neutrinos. In a broad archival cross-search between real-time high-energy track events from IceCube and the sample of classified  SNe IIn from the ZTF-BTS catalog, they searched for spatial and temporal coincidences within 100 days from the SN discovery time and the arrival of neutrinos with a 90\% containment region smaller than 30 sq. deg. The sizes of their final neutrino and SN-IIn samples are 138 and 163, respectively. They found, in addition to the \IC and \SN coincidence, another coincidence  between IceCube-231027A and SN2023syz. They reported a probability of finding two or more coincidences in the sample of $p(k\geq2)$ = 0.0067. \\

We repeated our statistical test with the same selection cuts of \cite{Lu2026} and the ZTF-BTS sample. Based on these selection criteria, we found the additional coincidence between SN2023syz and IceCube-231027A that was excluded in our original search of events reconstructed better than \IC. We found a probability of finding two or more random coincidences to be $p(k\geq2)$ = 0.006, consistent with the one found in \cite{Lu2026}. The $p$-value for $k\geq$ 2 is intriguing. However, we remark that this is very sensitive to the preselection criteria. Moreover, neutrinos with large uncertainties in their reconstructions are more likely to change their containment regions and arrival directions in updated releases, while previously identified multimessenger coincidences might disappear (e.g., \citealt{2025arXiv250706176Z}). \\

We note that when we repeated the cross-match with the same conditions of \cite{Lu2026}, but with the TNS catalog, we found the same two coincidences (see Table \ref{tab:all_coincidences}, where events IC210503A and IC240307A end up removed by the cut of 30 sq. deg on the neutrino 90\% area). However, after running the randomizations, we estimated a chance coincidence of observing a probability of finding two or more random coincidences to be $p(k\geq2)$ = 0.067. This is about ten times higher than the previous $p$-value and it is likely due to the larger sample of SN-IIn in the TNS in the same time frame investigated by \cite{Lu2026}; namely, with 381 instead of 169. The $p$-value is of the same order of magnitude of that computed in the test described in Sect. \ref{sec:broad_sample_SN}. We conclude that the fact that the $p$-value is so sensitive to the sample sizes weakens the neutrino-and-SN IIn connection. Moreover, it is sensitive to the cuts on the neutrino sample (e.g., error region size). Furthermore, the error regions of such events (especially those with large error contours) could be sensitive to the details of the reconstruction technique (e.g., \citealt{2025arXiv250706176Z}). Therefore, such error regions are likely to be uncertain and it is difficult to quantify this uncertainty in the analysis.
A larger number of neutrino events are also required to establish such a connection (see also \citealt{2025ApJ...978..133W}).

\section{Expected neutrino rate from \SN}\label{sec:neutrino_rate}
Although we have shown that the \IC/\SN coincidence is not at a statistically significant level (which is severely affected by the poor reconstruction of the neutrino event itself), we cannot simultaneously rule out the multimessenger association. Therefore, we can draw some simple considerations in the context of \SN being a neutrino source in light of the evidence of CSM interaction from our spectroscopic observations (see Sect. \ref{sec:spectroscopy}).

Unlike choked-jet scenarios \citep{2001PhRvL..87q1102M,2013PhRvL.111l1102M,2018ApJ...856..119H,2024A&A...690A.187Z} in which the neutrino production is efficient only in an optically thick environment at the early stages of the SN explosion, the proton acceleration and interaction with the CSM are efficient after the so-called shock breakout on timescales that can last up to several weeks (see e.g., \citealt{2012IAUS..279..274K}). Assuming the dense CSM has a wind density profile (and noting that its radial extent is much larger than the breakout radius\footnote{This is supported in our case by the shallow luminosity decline up to late times. For the correction of the profile and the resulting diffusion timescale for the finite dense CSM extent, see \cite{2026ApJ...998..247W}.}), the profile can be simply parameterized by only the breakout radius, $R_{bo}$, (defined at optical depth of $c/v_{bo}$) and the shock velocity at that radius, $v_{bo}$,

\begin{equation}\label{eq:density_profile}
    \rho(r) = \frac{c}{v_{bo}}\frac{m_{p}}{\sigma_T R_{bo}}(r/R_{bo})^{-2},
\end{equation}
where $c$ is the speed of light, $m_{p}$ the proton mass, and $\sigma_{T}$ the Thomson cross section (assuming the CSM is dominated by hydrogen). The rise time of the bolometric luminosity  \citep{2025arXiv250400098W} can be estimated as

\begin{equation}\label{eq:t_peak}
    \Delta t_{peak} \simeq 1.2 \frac{R_{bo}}{v_{bo}}log(c/v_{bo}).
\end{equation}
This assumes that the shock is driven in the CSM by an initially polytropic stellar envelope that expands into the CSM. In this case, the shock decelerates slowly as $v\propto r^{-0.1}$ (with only a weak dependence on the exact initial stellar density profile; e.g., \citealt{2025arXiv250400098W}). This weak deceleration proceeds as long as the swept CSM mass is lower than the ejecta mass, which is consistent with the obtained result of $M_{bo}\approx0.2M_\odot$, see below.

We can use Eq. (\ref{eq:t_peak}) to constrain $R_{bo}$. We estimated the explosion time in Sect. \ref{sec:photometry} to be $t_{\rm exp}=60723.21\pm1.22$ (MJD) and the peak time to be $t_{\rm peak}=60754.5\pm1.4$ (MJD). This gives $\Delta t_{\rm peak}\,\sim\,$31 days. The breakout velocity, $v_{bo}$, could not be measured directly from the spectrum. Photospheric velocities are typically measured by the blueshift absorption component of the broad H alpha P-cygni. However, there is no clear such absorption seen here (i.e., the feature to the blue of the wing is too narrow; therefore, it is likely noise). Therefore, we could not directly measure the shock velocity and assume a typical value of $v_{bo}=10^9~$cm/s (which is not expected to differ significantly). Plugging $\Delta t_{peak}$ and $v_{bo}$ in Eq. (\ref{eq:t_peak}), we get $\rm R_{bo} \simeq$ 6.6$\times$10$^{14}$ cm. We integrated the CSM density profile (Eq. \ref{eq:density_profile}) up to $R_{bo}$ and we find the CSM mass contained up to $R_{bo}$ is $M_{bo}\approx0.2M_\odot$. Therefore, we do not expect the shock to decelerate significantly during propagation of a few breakout radii (as the swept CSM mass is much lower than the ejecta mass, which is at the level of least a few $M_\odot$ for typical  SNe IIn).

Since the coverage of the light curve evolution is limited to the optical band, with only one observation in the NUV W1 band $\sim$40 days after the peak, we cannot accurately determine L$_{bol,p}$ from observations; however, the estimated value $L_{bol,p}$ = 2.3$\times$10$^{43}$ is compatible with the expectation from a CSM breakout scenario, $L_{bol,p} \simeq 10^{43}\Big( \frac{R_{bo}}{10^{14}\,\mathrm{cm}}\Big)\Big( \frac{v_{bo}}{10^{9}\,\mathrm{cm/s}}\Big)^{2}\,\,\,\, \mathrm{erg \,s^{-1}}$ \citep{2025arXiv250400098W}. In addition, the optical bands behavior also seems consistent with the expectation from the CSM breakout scenario, and the single X-ray upper limit obtained at 40 days past peak luminosity is consistent with the X-ray luminosity being smaller than 10\% of the bolometric luminosity at that epoch \citep{2025arXiv250400098W}.

The transition between the radiation-mediated shock (RMS) to the collisionless shock (CLS) occurs around 0.3 R$_{bo}$ \citep{2025arXiv250400098W}, while the proton acceleration and pion production are efficient approximately until the shock reaches $\sim$10 $\rm R_{bo}$ \citep{2012IAUS..279..274K}, with approximately constant neutrino luminosity during this period. The high-energy neutrino \IC was detected $\sim$63 days after the explosion (which corresponds to $\sim$7 $\rm R_{bo}$, including the weak shock deceleration), whereas the pion production should be efficient until $\sim$96 days from explosion ($\sim10 R_{bo}$).  The total nonthermal energy can be obtained following \cite{2012IAUS..279..274K}:

\begin{equation}\label{eq:E_nonthermal}
    E_{nt} \sim 10^{49}\Big( \frac{R_{bo}}{10^{14}\,\mathrm{cm}}\Big)^{2}\Big( \frac{v_{bo}}{10^{9}\,\mathrm{cm/s}}\Big)^{-1}\,\,\,\, \mathrm{erg},  
\end{equation}
where we assume that the accelerated protons carry a 10$\%$ fraction of the post-shock energy. We estimated a total E$_{nt}$ $\sim$ 4.3 $\times$ 10$^{50}$ erg. This is an order-of-magnitude estimation, which does not change significantly if the shock deceleration is included. Assuming an E$^{2}$ power-law spectrum for the protons with cutoffs at E$_{p,min}$ = 10$^{2}$ GeV and E$_{p,max}$ = 10$^{8}$ GeV, we calculated the expected neutrino rate for a detector with the effective area of the IceCube Astrotrack Bronze stream\footnote{\url{https://gcn.nasa.gov/missions/icecube}}, at the declination band of \SN. Integrating over a time window of 96 days, we find N$_{\nu_{\mu}}$ $\sim$ 1.5$\times$10$^{-3}$ expected muon neutrinos in the realtime stream event selection. This is consistent with the expected rates in other multimessenger associations with single high-energy neutrinos (e.g., \citealt{2021ApJ...912...54R}, \citealt{2025A&A...695A.266O}).

We repeated the same exercise for neutrinos starting from E$_{\nu,min}$ = 1 TeV, using the effective area used in the ten-year point source analysis, in the declination band of \SN \citep{2020PhRvL.124e1103A,2021arXiv210109836I}. We find an expected total number of N$_{\nu_{\mu}}$ $\sim$ 1.7$\times$10$^{-2}$ muon neutrinos within 96 days from the explosion. To estimate the detection rate, we need to multiply by the number of massive CSM SNe IIn, $\mathcal{O}(10^{2})$, observed during the IceCube operations, which gives $\mathcal{O}(0.1-1)$ expected detections. However, such detections will not be statistically significant. This estimate does not take into account the completeness of the catalogs used in the analysis. Finally, we briefly note that the nondetection in \textit{Fermi}-LAT data and the corresponding upper limits are consistent with the discussion on the gamma-ray optical depth and luminosity in \cite{2012IAUS..279..274K} and \cite{2025ApJ...978..133W}.


\section{Summary}\label{sec:summary}
We  present the first broad set of optical photometric and spectroscopic observations of the SN IIn \SN, found in coincidence with the IceCube realtime high-energy neutrino event \IC. The multimessenger association was first identified during the LAST neutrino follow-up program, although the SN was already discovered around 60 days before the neutrino detection. We present in this work for the first time high-resolution spectroscopic observations of \SN, which show clear evidence of CSM interaction at the time of the neutrino arrival. Although the optical spectrum indicates a promising environment for hadronic acceleration and neutrino production in a collisionless shock after the breakout, here we discuss how several factors challenge the significance of the multimessenger association. First, the poor reconstruction of the neutrino event weakens the spatial association. The $\sim$60-day delay affects the temporal coincidence, but it is consistent with the expected time frame in which neutrino production efficiency is high in these environments. We find that the spatial and temporal coincidence between \IC and \SN is consistent with the background hypothesis of being a chance coincidence, with the smallest $p$-value of $p(k\geq1)$ = 0.078 found against the ZTF-BTS catalog. We note that these $p$-values are sensitive to the size of the SNe and neutrino samples used for the statistical test. 

We additionally show how a broader search for neutrino and SN-IIn coincidences in the entire sample of IceCube alerts reveals two additional high-energy track-like events coincident with SNe  IIn  within a 100-day window since the SN discovery. However, randomizations of these samples reveal that the background hypothesis cannot be rejected for this emerging population of multimessenger coincidences.  

In conclusion,  while we cannot rule out the \SN-and-\IC association, we did compute the expected rate of neutrinos from \SN, assuming a typical scenario of proton acceleration and interaction in a collisionless shock. We found N$_{\nu_{\mu}}$ $\sim$ 10$^{-3}$ expected muon neutrinos within the time period of $\sim$96 days after the explosion in which \SN should, in principle, be an efficient neutrino source. This is consistent with the detection of a single high-energy neutrino event during this time. The expected rate is about a factor of 10 higher for neutrino energies as low as E$_{\nu,min}$ = 1 TeV. We encourage performing further dedicated point-source analysis of IceCube data from the direction of \SN to assess empirical constraints on these numbers. However, these estimates can vary significantly depending on the true characteristics of \SN, which could not be fully obtained from the available observations (e.g., see \citealt{2022ApJ...929..163P} for a detailed study on the SN+CSM model parameter space).

The multimessenger association between \IC and \SN does not provide any independent evidence of high-energy neutrino production in   SN IIn systems, but it can be considered among the population of possible neutrino sources identified by realtime multimessenger and multiwavelength observations.  In the near future, larger numbers of detected neutrinos, a continuous improvement of the realtime alerts, and synergies among the follow-up programs across the whole electromagnetic spectrum will be key to detecting an emerging population of sources that are currently below the detection level. These searches will also benefit from the advent of future larger detectors (e.g., IceCube-Gen2, \citealt{2021JPhG...48f0501A}) as well as new missions dedicated to the discovery and characterization of such transient events (e.g., ULTRASAT, \citealt{2024ApJ...964...74S}).  

\section*{Data availability}
All optical spectra from Sect. \ref{sec:spectroscopy} are available on WISeREP\footnote{WISeREP portal: \url{https://www.wiserep.org}}. Tables \ref{tab:LAST_photometry} and \ref{tab:ZTF_photometry} are only available in electronic form at the CDS via \url{https://cdsarc.
cds.unistra.fr/viz-bin/cat/J/A+A/708/A223}.

\begin{acknowledgements}
We thank the anonymous referee for useful comments on the manuscript. 
We thank B. Katz, E. Waxman, and B. Zackay for useful discussions. We thank P. M. Veres for useful comments to the manuscript and for triggering the Swift ToO.\\
S.G. is grateful for the support of the Koshland Family Foundation.\\
E.O.O. is grateful for the support of
grants from the 
Willner Family Leadership Institute,
André Deloro Institute,
Paul and Tina Gardner,
The Norman E Alexander Family M Foundation ULTRASAT Data Center Fund,
Israel Science Foundation,
Israeli Ministry of Science,
Minerva,
BSF, BSF-transformative, NSF-BSF,
Israel Council for Higher Education (VATAT),
Sagol Weizmann-MIT,
Yeda-Sela, and the
Rosa and Emilio Segrè Research Award.
This research is supported by the Israeli Council for Higher Education (CHE) via the Weizmann Data Science Research Center, and by a research grant from the Estate of Harry Schutzman.\\
S.A.S. Is grateful for the Zuckerman Scholars fellowship.\\
PC acknowledges the support from the Zhejiang provincial top-level research support program.\\
Based on observations made with the Liverpool Telescope operated on the island of La Palma by Liverpool John Moores University in the Spanish Observatorio del Roque de los Muchachos of the Instituto de Astrofisica de Canarias with financial support from the UK Science and Technology Facilities Council.\\
CMC receives funding from UKRI grant numbers ST/X005933/1 and ST/W001934/1.\\
This work made use of data supplied by the UK Swift Science Data Centre at the University of Leicester.\\
\end{acknowledgements}

%
%

\bibliographystyle{aa} 
\bibliography{biblio.bib} 

\end{document}